\documentclass[aps,prl,twocolumn,showpacs,superscriptaddress,floatfix]{revtex4-1} 

\usepackage{graphicx}  
\usepackage{amsmath}
\usepackage{bm}        
\usepackage{amssymb}   
\usepackage{gensymb}   
\usepackage{lipsum}

\usepackage{float}
\usepackage{color}

\begin{document}

\title{Disentangling the Effects of Curvature and Misorientation\\on the Shrinkage Behavior of Loop-Shaped Grain Boundaries}

\author{Fabrizio Camerin}%
\email{fabrizio.camerin@fkem1.lu.se}
\affiliation{ 
Soft Condensed Matter \& Biophysics, Debye Institute for Nanomaterials Science, Utrecht University, Princetonplein 1, 3584 CC Utrecht, The Netherlands
}%
\affiliation{International Institute for Sustainability with Knotted Chiral Meta Matter (WPI-SKCM$^2$), Hiroshima University, 1-3-1 Kagamiyama, Higashi-Hiroshima, Hiroshima 739-8526, Japan}
\affiliation{Physical Chemistry, Department of Chemistry, Lund University, SE-22100 Lund, Sweden}

\author{Susana Mar\'in-Aguilar}%
\email{susana.marinaguilar@uniroma1.it}
\affiliation{ 
Soft Condensed Matter \& Biophysics, Debye Institute for Nanomaterials Science, Utrecht University, Princetonplein 1, 3584 CC Utrecht, The Netherlands
}%
\affiliation{Department of Physics, Sapienza University of Rome, Piazzale Aldo Moro, 5, 00185, Roma, Italy}

\author{Tim Griffioen}%
\affiliation{ 
Soft Condensed Matter \& Biophysics, Debye Institute for Nanomaterials Science, Utrecht University, Princetonplein 1, 3584 CC Utrecht, The Netherlands
}%

\author{Mathieu G. Baltussen}%
\affiliation{ 
Institute for Molecules and Materials, Radboud University, Heyendaalseweg 135, 6525 AJ Nijmegen, The Netherlands
}%

\author{Roel P.A. Dullens}%
\affiliation{ 
Institute for Molecules and Materials, Radboud University, Heyendaalseweg 135, 6525 AJ Nijmegen, The Netherlands
}%

\author{Berend van der Meer}%
\affiliation{ 
Institute for Molecules and Materials, Radboud University, Heyendaalseweg 135, 6525 AJ Nijmegen, The Netherlands
}%
\affiliation{Physical Chemistry and Soft Matter, Wageningen University \& Research, Stippeneng 4, 6708 WE Wageningen, The Netherlands}

\author{Marjolein Dijkstra}%
\email{m.dijkstra@uu.nl}
\affiliation{ 
Soft Condensed Matter \& Biophysics, Debye Institute for Nanomaterials Science, Utrecht University, Princetonplein 1, 3584 CC Utrecht, The Netherlands
}%
\affiliation{International Institute for Sustainability with Knotted Chiral Meta Matter (WPI-SKCM$^2$), Hiroshima University, 1-3-1 Kagamiyama, Higashi-Hiroshima, Hiroshima 739-8526, Japan}

\date{\today}

\begin{abstract}
The material properties of polycrystals are strongly affected by the evolution and coarsening of their internal grain structures.
Yet, studying this process is challenging due to the complex interactions within grain boundary networks. Here, we systematically investigate the shrinkage of isolated loop-shaped grain boundaries in 2D colloidal crystals.
Unexpectedly, we find that shear coupling decreases with increasing grain misorientation, contrary to geometric predictions. This counterintuitive result is attributed to enhanced concurrent sliding driven by the annihilation of dislocations. 
Furthermore, by focusing on the evolution of the grain size, we reveal a transition in shrinkage kinetics between small and large loop sizes, offering an explanation for previously observed discrepancies in grain boundary mobility. These findings reveal a more intricate dependence of grain boundary behavior on curvature and misorientation than previously reported, offering new insights into polycrystal coarsening dynamics.  
\end{abstract}

\maketitle

Throughout the history of forging tools, blacksmiths have employed a diverse range of techniques to enhance the material properties of metals, such as their hardness and strength, through the controlled application of heat. These methods rely on altering the microstructure, particularly by modifying the average size of the crystal domains within the material~\cite{howe1997interfaces,hall1951deformation,petch1953cleavage}. Each crystal domain, known as a grain, has a distinct crystallographic orientation, and is separated from adjacent domains by an interface, referred to as a grain boundary (GB). On the atomic scale, the motion of GBs is facilitated by the displacement of atoms across the GB from one grain to another. Besides the fundamental interest aimed at unraveling the microscopic mechanisms at play during this process~\cite{cahn2004unified,cahn2006coupling,trautt2012grain,rajabzadeh2013elementary,zhu2019situ,wei2021direct}, understanding the motion of GBs is crucial in various applications, ranging from small-scale to industrial~\cite{cantwell2020grain,gottstein2009grain,han2018grain,karbasian2010review, quirk2024grain}. 
However, the highly interconnected nature of GB networks in real materials implies that the motion of an individual GB is significantly influenced by forces arising from interactions with neighboring GBs~\cite{thomas2017reconciling,zhang2017equation,zhang2021equation,rohrer2023grain}. This interdependence complicates efforts to determine whether its dynamics is driven by the intrinsic properties of the GB or by external interactions.

Grain boundary loops (GBLs) stand out as a fascinating exception since these self-contained structures consist of a fully isolated, single GB that is embedded within a large domain of a two-dimensional crystal~\cite{kurasch2012atom,gong2016situ,lavergne2018dislocation,lavergne2019shrinkage,chaffee2024hexagonal}.  As a result, this GB geometry is characterized by two key parameters: its radius (or curvature) and the misorientation angle, which is the difference in crystallographic orientation between the enclosed domain and the surrounding bulk crystal. A defining characteristic of GBLs is their inherent instability, causing them to shrink over time, and eventually annihilate completely. Shrinkage can occur through three distinct mechanisms~\cite{cahn2004unified,cahn2006coupling,trautt2012grain,lavergne2019shrinkage}.
In the simplest case, it occurs \textit{purely} through {\em curvature-driven} migration~\cite{mullins1956two}, with the GBL moving toward its center under a pressure proportional to its curvature. In this scenario, no grain rotation occurs, and the misorientation remains constant throughout the process~\cite{cahn2004unified,trautt2012grain}.
More commonly, however, shrinkage is accompanied by grain rotation. This can occur either toward lower misorientations to minimize interfacial (free) energy, a process known as {\em sliding},
or toward higher misorientations through {\em coupled} GB migration, where normal and tangential motions are interconnected~\cite{cahn2004unified,cahn2006coupling,cahn2006duality,trautt2012grain}. At the particle level, this coupled motion ensures the continuity of particle rows at the boundary, 
leading to the rotation of the enclosed grain relative to bulk. 
We note that sliding and coupled migration can also occur concurrently, potentially balancing each other and {\em effectively} resulting in curvature-driven GB migration without significant grain rotation.
A schematic illustrating these three shrinking mechanisms can be found in the Supplemental Material (SM)~\cite{sm}.

\begin{figure*}
\begin{center}
\includegraphics[width=0.95\linewidth]{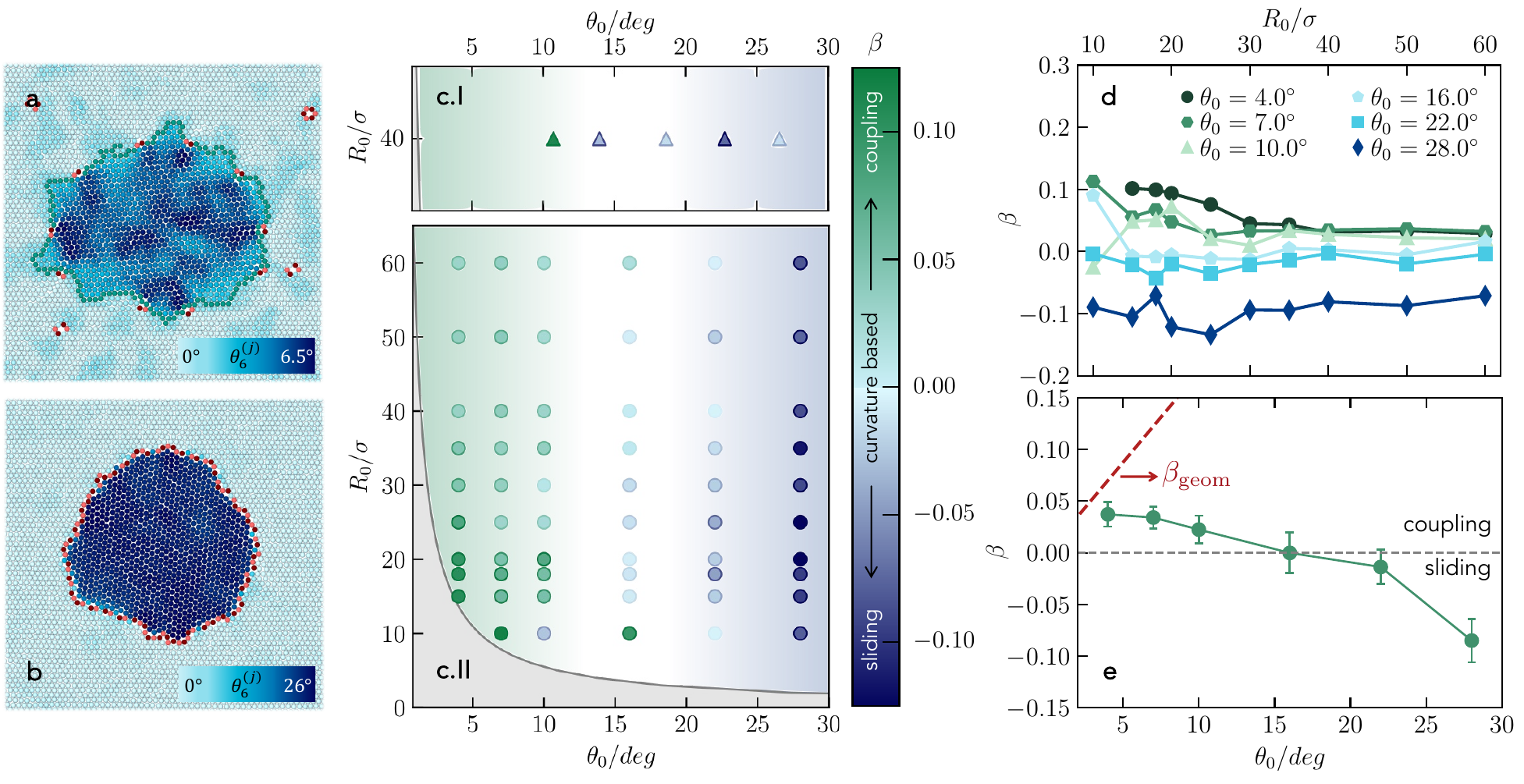}
\end{center}
\vspace{-0.5cm}
\caption{
\textbf{State diagram.} (a,b) Representative simulation snapshots for systems with $R_0=50\sigma$ and $\theta_0=4^\circ$ and $28^\circ$, respectively. Particles are colored based on $\theta_6^{(j)}$, as indicated by the color bar in each panel. Dislocations are denoted as pink/red particle pairs. In (a) the GBL is highlighted in green for visual clarity, while in (b), it is enclosed by dislocations. (c) Dominant GBL shrinking mechanism for varying $R_0$ and $\theta_0$ in (I) experiments and (II) simulations, according to $\beta$, as denoted by the color bar. The gray area corresponds to $R_0 \theta_0 < 3 a/\pi$. (d,e) $\beta$ as a function of $R_0$ and $\theta_0$, respectively. The red dashed line in (e) corresponds to $\beta_\mathrm{geom}=2\tan(\theta/2)$.}
\label{fig:statediagram}
\end{figure*}

While GBLs can spontaneously form in engineered materials~\cite{cockayne2011grain,yazyev2014polycrystalline}, direct studies remain particularly challenging due to their unstable nature.
Colloidal particles have emerged as a convenient platform for studying GBs~\cite{swinkels2023visualizing,menath2023acoustic,lobmeyer2022grain,van2016fabricating,liao2018grain,gokhale2013grain,nagamanasa2011confined,camerin2024depletion,maire2016imaging,irvine2013dislocation,lavergne2017anomalous} and specifically GBLs, which can be created by manipulating crystalline lattices of colloidal particles using optical tweezing~\cite{lavergne2018dislocation,lavergne2019shrinkage,chaffee2024hexagonal}.
In this way, elastic relaxation~\cite{lavergne2018dislocation} and grain splitting~\cite{barth2021grain} have been demonstrated as novel mechanisms for coarsening at small grain sizes.
However, regarding the shrinkage kinetics of colloidal GBLs conflicting results have been reported on the effect of the misorientation. More specifically, experimental work reported a shrinkage rate that diverges at small misorientation angles~\cite{lavergne2018dislocation}, while a numerical study did not observe this behavior~\cite{guo2019competing}. 
Furthermore, although  experimental studies have observed both coupling and sliding behaviors in colloidal GBLs~\cite{lavergne2018dislocation, lavergne2019shrinkage}, these findings have predominantly been qualitative and lack systematic analysis. The absence of systematic and quantitative data leaves crucial questions unresolved, including  the factors that determine the dominant mechanism, their impact  on shrinkage dynamics, and the specific roles of misorientation and loop size.

In this Letter, we utilize computer simulations to extensively investigate GBLs in 2D colloidal crystals. We construct a detailed state diagram that reveals the dominant shrinkage mechanisms as a function of size and misorientation, showing qualitative agreement with experimental observations. Unexpectedly, we observe a decrease in shear coupling with increasing misorientation, defying established geometric models. 
We show that this behavior is due to the interplay of coupling and enhanced concurrent sliding, a process we microscopically link to dislocation dynamics. Furthermore, we uncover a transition in shrinkage kinetics, where small and large GBLs exhibit distinct mobility regimes. These findings provide fresh insights for understanding the fundamental mechanisms governing the evolution of grain structures.

\begin{figure*}
\begin{center}
 \includegraphics[width=\linewidth]{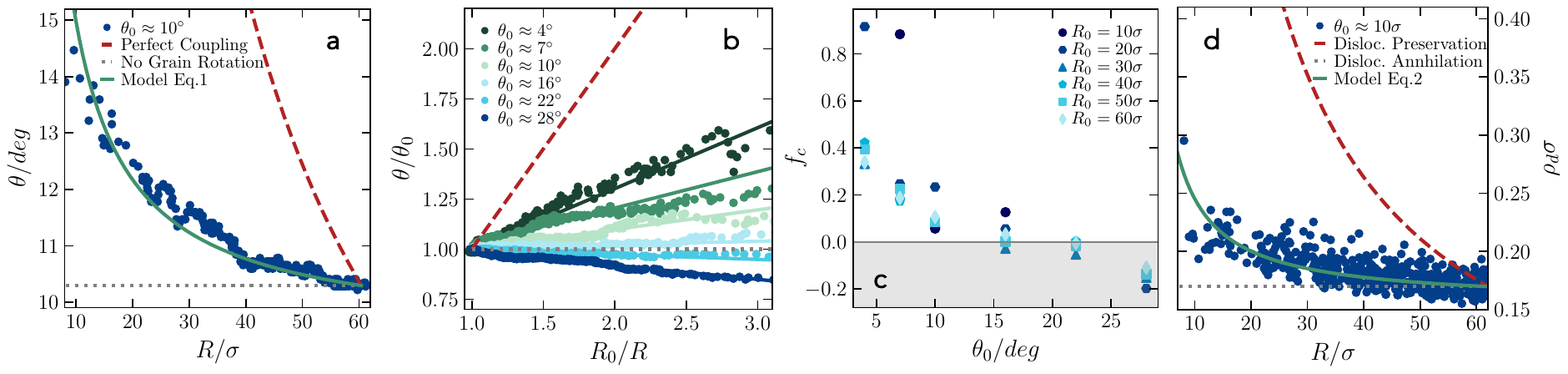}
 \end{center}
 \vspace{-0.5cm}
 \caption{\textbf{Imperfect coupling.} (a) Shrinkage trajectory showing $\theta$ as a function of $R$ for $R_0=60\sigma$ and $\theta_0=10^{\circ}$, with  the green line corresponding to the fit to Eq.~\ref{eq:lincomb}. (b) Shrinkage trajectories plotted as $\theta/\theta_0$ versus $R_0/R
 $, for $R_0=60\sigma$ and varying $\theta_0$. Full lines correspond to fits to Eq.~\ref{eq:lincomb}. (c) Extracted values of $f_c$ as a function of $\theta_0$ for varying $R_0$. The grey area denotes $f_c<0$. (d) $\rho_d$ as a function of $R$ for $R_0=60\sigma$ and $\theta_0\approx 10^{\circ}$, with  the green line being the fit to Eq.~\ref{eq:rho_d}. In (a, b, d), the dashed red line represents perfect coupling while the gray dotted line corresponds to shrinkage without grain rotation.
}
\label{fig:partial_coupling}
 \end{figure*}

We perform Brownian dynamics simulations of a two-dimensional hexagonal crystalline monolayer consisting of $N$ hard-sphere-like particles with a diameter $\sigma$, which is taken as the unit of length, at a packing fraction $\eta=0.68$ (see SM~\cite{sm}).
We create GBLs by mimicking the experimental procedure that uses an optical vortex to rotate a circular patch within a large single crystal~\cite{lavergne2018dislocation, lavergne2019shrinkage}. To this end, we apply a torque-like force to all particles within a radius $R_0$ until the desired misorientation $\theta_0$ is reached. Subsequently, we turn off the rotation and allow our bicrystalline system to evolve, free from external forces, until the grain completely vanishes. 
We systematically investigate the shrinkage process by examining multiple realizations of GBLs  for a range of initial radii $10 \leq R_0/\sigma \leq 60$, and initial misorientation angles $4^\circ \leq \theta_0 \leq30^\circ$. Representative simulation snapshots are shown in Fig.~\ref{fig:statediagram}(a,b) for two different $\theta_0$.
Additionally, we compare our simulation results with a limited set of experimental data, in which we make use of a two-dimensional colloidal model system with hard-sphere interactions~\cite{thorneywork2017two} and employ the optical vortex method to generate GBLs~\cite{lavergne2018dislocation,lavergne2019shrinkage}. The shrinkage of the GBL is monitored at the single-particle level using bright-field microscopy. 
During the shrinkage of the GBL, we track the misorientation $\theta(t)$ and  grain radius $R(t)$ over time in both simulations and experiments. For each particle $j$, we monitor its position $\bm{r}_j$, local orientation $\theta_6^{(j)}$ (blue shaded particles in Fig.~\ref{fig:statediagram}(a,b)), and number of nearest neighbors $\mathcal{N}_{NN}^{(j)}$. Dislocations are identified as particle pairs with $\mathcal{N}_{NN}^{(j)} \neq 6$ (pink/red particles in Fig.~\ref{fig:statediagram}(a,b)).  
More details on the simulations and experiments are provided in the SM~\cite{sm}.

To study the dependence of the initial grain size and misorientation on the resulting shrinking mechanisms, 
we begin by computing the shear coupling factor, defined as $\beta=-d\theta/d\ln{(R/\sigma)}$~\cite{cahn2004unified,cahn2006coupling,trautt2012grain} (see SM~\cite{sm}). If $\beta > 0$, the grain couples to the bulk crystal while shrinking. Conversely, when $\beta < 0$, the grain slides along the boundary towards lower misorientation angles. Finally, when $\beta \approx 0$, the grain shrinks primarily due to its own curvature.  
We summarize the shrinkage behavior in the state diagram in Fig.~\ref{fig:statediagram}c, corroborated by experimental data at varying $\theta_0$ in panel I. The state diagram is colored  according to $\beta$. Since GBLs can only exist in hexagonal lattices above the critical loop threshold  $R_0 \theta_0 = 3a/\pi$~\cite{lavergne2018dislocation} with $a$ the lattice spacing, we shade this region in gray in Fig.~\ref{fig:statediagram}c. 
We observe that for small initial misorientations $\theta_0 \lesssim 10\degree$, grain coupling dominates the state diagram, as supported by consistently positive values of $\beta$.
For misorientation angles $10^\circ < \theta_0 < 22^\circ$,  a crossover region emerges where curvature-driven shrinkage predominantly  restores  the crystal lattice. While coupling and sliding may still occur,  their effects largely counterbalance, leading to  minimal net grain rotation. At high misorientation angles, we find $\beta < 0$, indicating that grain sliding dominates the shrinkage mechanism. 
Notably, our analysis reveals no dependence of $\beta$ on the initial grain size across the entire range of misorientations investigated (Fig.~\ref{fig:statediagram}d), with $\beta$ remaining constant regardless of  initial grain size. 
Minor deviations from this constant value are likely due to statistical fluctuations in the measurements and challenges in accurately tracking grains with very small $R$ and $\theta$.

A quantitative comparison with the idealized shear coupling behavior can be made by examining the relationship between $\beta$ and $\theta_0$. For a grain perfectly coupled to the bulk crystal, geometry dictates that the shear coupling factor is given by $\beta_{\mathrm{geom}}(\theta) = 2 \tan(\theta/2)$~\cite{cahn2004unified,cahn2006coupling,cahn2006duality, trautt2012grain}.
The comparison is shown in Fig.~\ref{fig:statediagram}(e), with  the red dashed line indicating the geometrical prediction.
Remarkably, the measured $\beta$ is always lower than $\beta_{\mathrm{geom}}$, indicating  imperfect coupling between the rotated grain and the bulk crystal. This suggests that, while coupled grain boundary migration dominates at low misorientation, significant grain sliding occurs simultaneously.

\begin{figure}[t!]
\begin{center}
\includegraphics[width=0.95\linewidth]{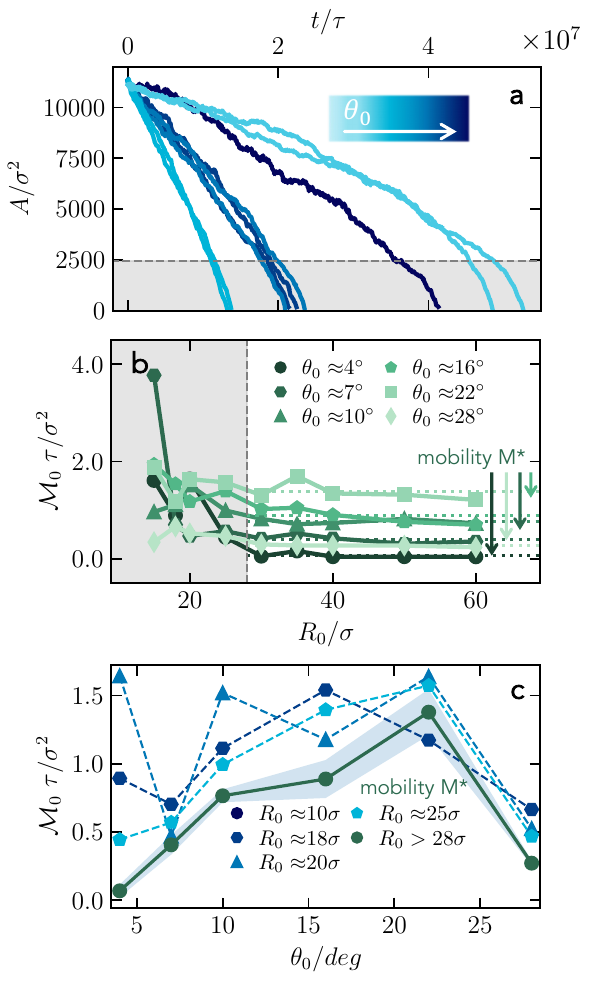}\end{center}
\vspace{-0.5cm}
\caption{
\textbf{Size-dependent shrinkage kinetics.} (a) Grain area $A(t)$ as a function of time $t$ for $R_0=60\sigma$ and varying $\theta_0$. (b,c) GBL initial shrinkage rate $\mathcal{M}_0$ and reduced mobility $M^\ast$ as a function of $R_0$ and $\theta_0$, respectively. In (a,b), the empirical threshold distinguishing small and large loops is denoted by the grey dashed line.
}
\label{fig:mobilities}
\end{figure}

Intriguingly, we observe that at low misorientations, where coupling dominates, $\beta$ decreases with increasing $\theta_0$, contrary to the trend expected by the geometric relation.
To explore this unexpected behavior, we plot the shrinkage trajectory of the GBL as $\theta$ versus $R$ for a representative state point in Fig.~\ref{fig:partial_coupling}a.
Next, we compare this to two limiting cases. The first scenario represents an idealized case of perfectly coupled GB migration, whose evolutions is given by  $\theta(R(t)) \approx R_0\theta_0/R(t)$ (red dashed line)~\cite{cahn2004unified,trautt2012grain,lavergne2019shrinkage}. 
The second scenario corresponds to 
a grain shrinking without any grain rotation, $\theta(R(t)) = \theta_0$ (dotted gray line). This could arise either from purely curvature-driven shrinkage or when sliding and coupling perfectly counterbalance each other.
As shown in Fig.~\ref{fig:partial_coupling}a, the GBL is far from perfectly coupled. Instead, the relatively weak increase in misorientation during the shrinkage suggests that coupling is accompanied by simultaneous sliding. 
Interestingly, we find that the shrinkage trajectory can be well described by a linear combination of the two limiting behaviors, as
\begin{equation}
    \theta(R) = f_{c} \theta_0 R_0/R +(1-f_{c})\theta_0 
    \label{eq:lincomb}
\end{equation}
where the extent of coupling is directly quantified through $f_{c}$. Namely, when $f_{c}=1$, the GBL is perfectly coupled with no sliding. Conversely, when $f_{c}=0$, sliding and coupling perfectly balance each other. Hence, by fitting the shrinkage trajectories of the GBL to Eq.~\ref{eq:lincomb}, we can extract $f_{c}$ for a range of $\theta$, thus measuring the relative contributions of coupling and sliding during the shrinkage process.
This is most conveniently done by re-plotting the shrinkage trajectories as $\theta(t)/\theta_0$ vs $R_0/R(t)$ and fitting with straight lines (see Fig.~\ref{fig:partial_coupling}b). The extracted values of $f_c$ are shown in Fig.~\ref{fig:partial_coupling}c.
Interestingly, we find that $f_{c}$ decreases rapidly with increasing misorientation, implying that concurrent sliding becomes more pronounced. In other words, the probability of the grain boundary migrating through displacement events characteristic of perfectly coupled GB motion diminishes significantly. As a result the effective shear coupling factor $\beta$ is influenced by this concurrent sliding through the prefactor $f_{c}(\theta)$, such that $\beta (\theta) = f_{c}(\theta) \beta_{geom}(\theta)$. This relationship accounts for the surprising observation that $\beta$ decreases as $\theta$ increases (see Fig.~\ref{fig:statediagram}e).
We remark that this analysis is valid only when $0 \leq f_c \leq 1$, as values outside this range would invoke an unphysical combination of shrinkage scenarios. As a result, it cannot account for cases where sliding dominates (shaded area in Fig.~\ref{fig:partial_coupling}c), and is only applicable at low misorientations, where there is net coupling. It is surprising, however, that a linear relationship persists even when $f_c$ is outside this physically meaningful range (Fig.~\ref{fig:partial_coupling}b).

To shed light on the microscopic origin of $f_{c}$, we recall that the ideal coupling scenario corresponds to the preserved continuity of crystal rows at the GB. This also corresponds to shrinkage occurring without dislocation reactions~\cite{cahn2004unified}. 
However, the loop-shaped geometry of our GBs makes dislocation annihilation unavoidable, since they are bound to encounter each other during the shrinkage process.
Therefore, we infer that the reason for $f_{c}< 1$ stems from these annihilation events.
To test this hypothesis, we track the number density of dislocations $\rho_d =n_d/L_{GBL}$, with $n_d$ the number of dislocations and $L_{GBL}$ the GBL perimeter. 
In full analogy to the treatment we presented earlier for the shrinking trajectories (see Fig.~\ref{fig:partial_coupling}a), it can be described by a linear combination of full dislocation preservation (perfect coupling) and no dislocation preservation (no grain rotation), which can be expressed as (see SM~\cite{sm}): 
\begin{equation}
       \rho_d(R) = f_p \rho_{d,0}R_0/R +(1-f_p) \rho_{d,0}, 
       \label{eq:rho_d}
\end{equation}
where $\rho_{d,0}$ is the initial dislocation density and $f_p$ is the fraction of dislocations preserved during the shrinkage process. 
Remarkably, we find that the dislocation density $\rho_d(R)$ is accurately described by imposing $f_p=f_c$ (green line in Fig.~\ref{fig:partial_coupling}d). 
To further validate this, we obtain $f_p$ for each state point by treating it as a fitting parameter and compare it to $f_c$ as extracted using Eq.~\ref{eq:lincomb}. Remarkably, a strong correlation $f_p \approx f_c$ is observed (see SM~\cite{sm}), thus establishing a clear connection between the macroscopic coupling behavior and the underlying microscopic dislocation dynamics. This equivalence is not just observational; the connection between the microscopic preservation of dislocations and the extent of coupling $(f_p = f_c)$ is also mathematically evident, as Eq.~\ref{eq:lincomb} and \ref{eq:rho_d} become identical when we approximate $\rho_d a\approx \theta$~\cite{howe1997interfaces}.

Having assessed the mechanism of GBL shrinkage, we now quantitatively analyze their shrinkage kinetics.
We begin by examining grains with identical $R_0$ but varying $\theta_0$, by plotting the grain area $A$ as a function of time $t/\tau$, with $\tau$ the unit of time in simulations (see SM~\cite{sm}), as shown in Fig.~\ref{fig:mobilities}a.  
When the GBLs are large, their area simply decreases linearly in time, consistent with capillary-driven grain shrinkage described by $\frac{dA}{dt}=-2 \pi M^\ast\label{eq:shr}$, where $M^\ast$ is commonly referred to as the reduced GB mobility~\cite{mullins1956two,lavergne2019shrinkage}. 
However, a deviation from this linear behavior arises once $A\lesssim 2500\sigma^2$, corresponding to loop sizes comprised of a few thousand particles typically studied in previous works~\cite{lavergne2018dislocation,lavergne2019shrinkage,guo2019competing}. 
A detailed window-averaged shrinkage rate analysis, presented in the SM~\cite{sm}, further demonstrates this deviation and clearly reveals a significant increase in the shrinkage rate at small loop sizes.

To identify the critical GBL size associated with this transition, we extract the initial shrinkage rate $\mathcal{M}_0=-\langle dA/dt\rangle /2 \pi $ by analyzing the area decay during the \emph{initial} stage of GBL shrinkage (see SM~\cite{sm}).
This approach enables us to establish the GBL size regime where shrinkage occurs linearly and is not affected by the non-linear kinetics at small loop sizes.
In Fig.~\ref{fig:mobilities}b we show the relationship between $\mathcal{M}_0$ and the initial radius $R_0$ for different initial misorientations $\theta_0$. A clear large-to-small loop transition in the shrinkage rate is indeed observed around $R_0 \approx 28\sigma$ (shaded area). Beyond this threshold, $\mathcal{M}_0$ remains  constant with respect to $R_0$, indicating that only in this regime does the grain shrink at a well-defined, constant rate $\mathcal{M}= M^\ast$ (dashed lines). We note this large-to-small GBL transition is reminiscent of earlier observations in half-loop bicrystals, where high GB curvature led to non-linear shrinkage dynamics, and a constant shrinkage rate was only observed for lower curvatures~\cite{upmanyu1998atomistic,upmanyu1999misorientation,zhang2005curvature}.

Furthermore, we study the effect of the initial misorientation on the shrinkage kinetics, by now plotting $\mathcal{M}_0$ as a function of $\theta_0$ in Fig.~\ref{fig:mobilities}c. Consistently, for $R_0 > 28\sigma$, this corresponds to $\mathcal{M}_0=M^\ast$,  indicated by the horizontal dotted lines in Fig.~\ref{fig:mobilities}(b). 
The shaded area shows the (statistical) variation of $M^\ast$. Notably, we observe that the reduced GB mobility $M^\ast$ exhibits a non-monotonic dependence on $\theta_0$. In particular, it reaches a maximum at approximately $\theta_0 \approx 22^\circ$, which is in qualitative agreement with previous numerical work~\cite{upmanyu2006simultaneous}.
Crucially, our results help resolve apparent contradictions in the reported misorientation dependence of $M^*$. Specifically, while experimental studies have observed a divergence of $M^*$ at small misorientation angles~\cite{lavergne2018dislocation}, this behavior was not observed in a  numerical study~\cite{guo2019competing}.
These discrepancies likely stem from both studies being conducted within the small-loop regime.
We hypothesize that in the large-loop regime, capillary pressure-driven dislocation motion is the main driver for GB motion, with neighboring dislocations primarily migrating parallel to one another. This leads to a well-defined GB mobility. In contrast, in the small-loop regime, neighboring dislocations have differing orientations with intersecting glide planes. As a result, direct interactions between dislocations become increasingly important and, even subtle variations in dislocation structure, can significantly affect the shrinkage kinetics.
This hypothesis is further supported by the finding that for small GBLs different grain boundary construction methods can lead to different mobility values~\cite{guo2019competing}.

In conclusion, we conducted a systematic investigation of the shrinkage of GBLs across a wide range of initial radii and misorientation angles. Our findings challenge existing theoretical predictions by demonstrating a decrease in shear coupling with increasing misorientation. This counterintuitive behavior arises from concurrent sliding, which we link microscopically to dislocation annihilation. Furthermore, we identified a transition in shrinkage kinetics between small and large loops, offering a potential resolution to previously reported discrepancies regarding the misorientation dependence of the GB mobility. 
Ultimately, our results illuminate the complex interplay of curvature and misorientation in dictating GB behavior in polycrystals, thereby serving as a crucial stepping stone for designing materials with tailored properties.

\medskip
\begin{acknowledgements}
F.C., S.M.-A., T.G. and M.D. acknowledge funding from the European Research Council (ERC) under the European Union's Horizon 2020 research and innovation program (Grant agreement No. ERC-2019-ADG 884902, SoftML). F.C. and M.D. acknowledge funding from the World Premiere International (WPI) Research Center Initiative of the Japanese Ministry of Education, Culture, Sports, Science and Technology (MEXT). B.vdM. acknowledges funding from the Netherlands Organization for Scientific Research (NWO) through a Rubicon grant (Rubicon Grant No. 019.191EN.011) and Veni grant (Veni Grant No. VI.Veni.212.138). R.P.A.D. acknowledges the ERC (Consolidator Grant No. 724834-OMCIDC).
\\\\
\indent F.C., S.M.-A. and T.G. contributed equally to this work. Author contributions are defined based on CRediT (Contributor Roles Taxonomy). Conceptualization: F.C., S.M.-A., R.P.A.D., B.vdM., M.D.; Formal analysis: F.C., S.M.-A., T.G., B.vdM.; Funding acquisition: R.P.A.D., B.vdM., M.D.; Investigation: F.C., S.M.-A., T.G., M.G.B., B.vdM., R.P.A.D., M.D.; Methodology: F.C., S.M.-A., T.G., B.vdM.; Project administration: R.P.A.D., M.D.; Software: T.G.; Supervision: F.C., S.M.-A., R.P.A.D., B.vdM., M.D.; Validation:  F.C., S.M.-A., T.G., B.vdM.; Visualization: F.C., S.M.-A.,  T.G.; Writing – original draft: F.C., S.M.-A., T.G., B.vdM., M.D.; Writing – review and editing: F.C., S.M.-A., T.G., R.P.A.D., B.vdM., M.D..
\end{acknowledgements}

\clearpage
\newpage
\onecolumngrid
\begin{center}
\large
\textbf{Disentangling the Effects of Curvature and Misorientation\\on the Shrinkage Behavior of Loop-Shaped Grain Boundaries\\ \bigskip Supplementary Material}

\normalsize
\bigskip
Fabrizio Camerin\textsuperscript{ 1,2,3}, Susana Marín-Aguilar\textsuperscript{ 1,4}, Tim Griffioen\textsuperscript{ 1}, \\Mathieu G.
Baltussen\textsuperscript{ 5}, Roel P.A. Dullens\textsuperscript{ 5}, Berend van der Meer\textsuperscript{ 5,6}, Marjolein Dijkstra\textsuperscript{ 1,2}\\
\medskip
\small
\textit{%
\textsuperscript{1}Soft Condensed Matter \& Biophysics, Debye Institute for Nanomaterials Science, Utrecht University, Princetonplein 1, 3584 CC Utrecht, The Netherlands\\
\textsuperscript{2}International Institute for Sustainability with Knotted Chiral Meta Matter (WPI-SKCM$^2$), Hiroshima University, 1-3-1 Kagamiyama, Higashi-Hiroshima, Hiroshima 739-8526, Japan\\
\textsuperscript{3}Physical Chemistry, Department of Chemistry, Lund University, SE-22100 Lund, Sweden\\
\textsuperscript{4}Department of Physics, Sapienza University of Rome, Piazzale Aldo Moro, 5, 00185, Roma, Italy\\
\textsuperscript{5}Institute for Molecules and Materials, Radboud University, Heyendaalseweg 135, 6525 AJ Nijmegen, The Netherlands\\
\textsuperscript{6}Physical Chemistry and Soft Matter, Wageningen University \& Research, Stippeneng 4, 6708 WE Wageningen, The Netherlands
}

\end{center}

\normalsize
\bigskip
\renewcommand{\theequation}{S\arabic{equation}}\setcounter{equation}{0}
\renewcommand{\thefigure}{S\arabic{figure}}\setcounter{figure}{0}
\renewcommand{\thetable}{S\arabic{table}}\setcounter{table}{0}

\twocolumngrid
\section{Simulation details}

\subsection{System preparation}

\begin{figure*}[t!]
\begin{center}
\includegraphics[width=1\linewidth]{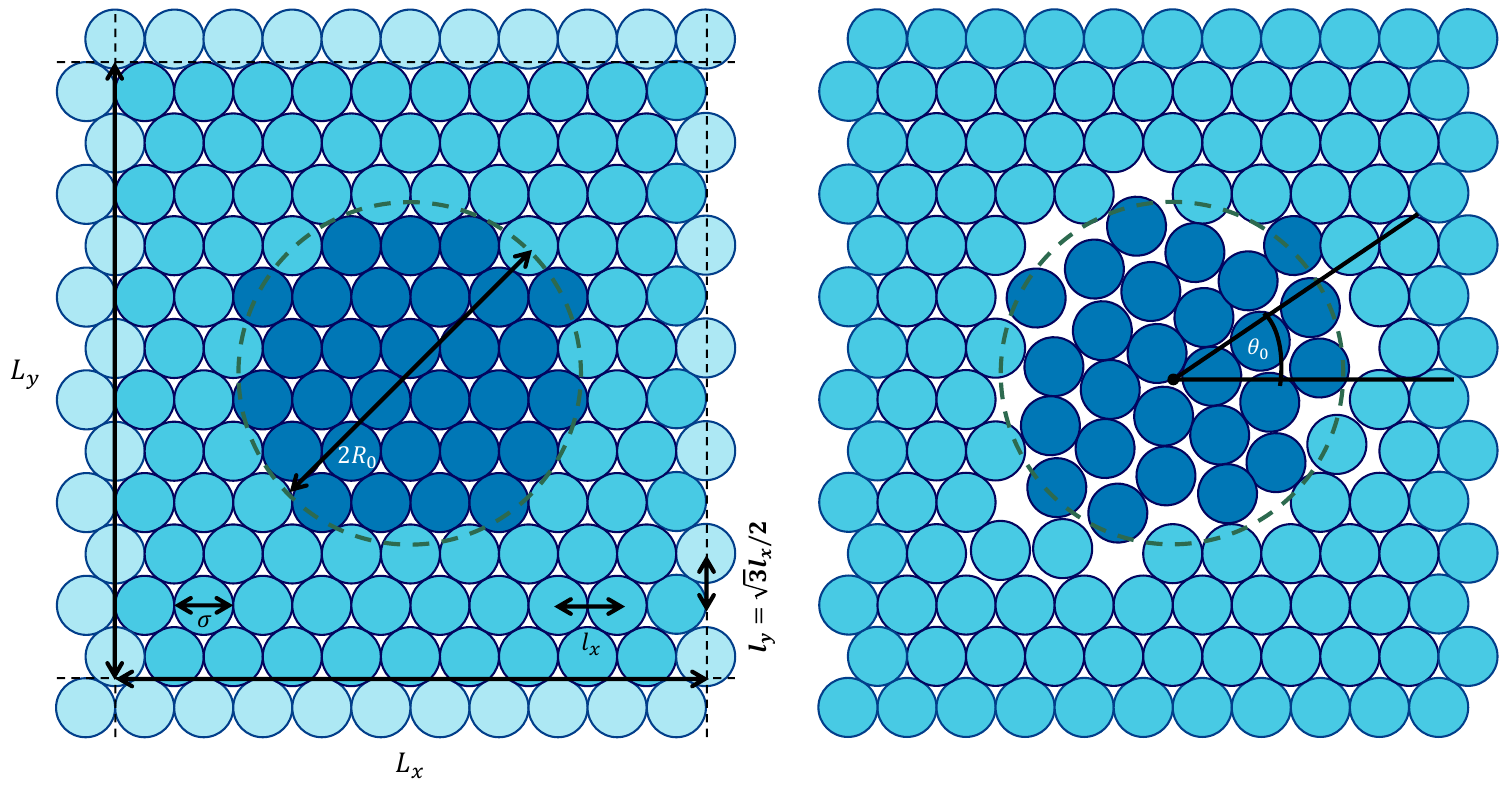}
\end{center}
\vspace{-0.5cm}
\caption{(a) Schematic representation of the simulation setup. Circles represent colloidal particles with diameter $\sigma$. The distance between particles in a row is $l_x$, while the distance between consecutive rows is $l_y$. The simulation box with sides $L_x$ and $L_y$ is depicted by black dashed lines and light blue circles represent periodic images. For a homogeneous crystal at the periodic boundaries, an even number of rows is required. After equilibrating the system, a rotational force is applied to all particles (dark blue) positioned within a circular patch of radius $R_0$ (green dashed line). (b) Schematics of the system after the torque has been applied to particles. The grain retains an overall initial misorientation angle $\theta_0$ compared to the bulk crystal.}
\label{fig:simulation_box_schematic}
\end{figure*}

We consider a two-dimensional hexagonal crystalline monolayer consisting of hard-sphere-like particles with a diameter $\sigma$ and a mass $m$. 
The particles are restricted to move exclusively within two dimensions. To create  a two-dimensional bicrystalline system with a grain boundary loop (GBL), we employ the following procedure. We first compute the horizontal length of the hexagonal unit cell $l_x/\sigma=\sqrt{\eta_{HS}/\eta}$ for a specified packing fraction $\eta=\pi \sigma^2 N/ 4A$, where $N$ denotes the number of particles, $A$ the  area of the system, and $\eta_{HS}$ the maximum  hard-sphere packing fraction of a two-dimensional monolayer.  Subsequently, we  determine the dimensions of the simulation box required to accommodate a GBL with a specified initial radius $R_0$. For a GBL with an initial radius $R_0$, we create a simulation box with an approximate length of $\tilde{L} = 2R_0 + 32\sigma$. The additional $32\sigma$ is included to ensure that the GBL does not interact  with any of its periodic images.
Finally, we create  rows and columns of particles within the crystalline monolayer. 
We determine the effective horizontal box length $L_x$ by rounding $\tilde{L}$ up to the nearest  integer multiple of the horizontal length of the unit cell 
\begin{equation}
    L_x = l_x N_x = l_x \left\lceil \frac{\tilde{L}}{l_x} \right\rceil,
\end{equation}
where $N_x$ represents the number of particles in a row, and the  symbol $\lceil$ denotes the ceiling function. In a hexagonal unit cell, successive rows exhibit a horizontal offset of 
$\l_x$, and the distance between crystal rows is given by  
$l_y = \sqrt{3}l_x/2$. 
To keep the simulation box as square as possible, we determine the number of crystal rows as $N_y = L_x / l_y$ and round it to the nearest even integer multiple of $l_y$. Fig.~\ref{fig:simulation_box_schematic} shows a schematic representation of the final simulation box.

We equilibrate the system  for a duration of $t = 2 \times 10^4 \delta t$, where $\delta t = 0.001 \tau$ represents the integration time step, and $\tau=\sqrt{m \sigma ^2 / \epsilon}$ is the unit of time in our simulations, with $\epsilon$ determining the energy scale. 
Subsequently, we create a GBL by applying a torque-like force $\bm{F}_{rot}(\bm{r}_i)$ to all particles $i$ within a circular patch of radius  $R_0$ centered at the origin  of the simulation box, until the average misorientation angle of the grain reaches a specified value $\theta_0$.  We then switch off the torque-like force, set the time $t=0$, and let the system evolve until the grain has vanished (see below).  The force scales linearly with the distance to the center to preserve the local crystalline  order of the circular patch and reads 
\begin{align}
\bm{F}_{rot}(\bm{r}_i) = -\tilde{F}(\bm{r}_i \times \bm{\hat{z}}),
\end{align}
where $\tilde{F} = 0.2~k_B T / \sigma^2$ represents the strength of the rotational force,  $k_B$  the Boltzmann constant,  $T$ the temperature,  $\bm{r}_i$  the position of particle $i$, and $\bm{\hat{z}}$  the unit vector pointing in the z-direction out of the plane of the crystal.

\subsection{Simulations}
\begin{figure}[t!]
\begin{center}
\includegraphics[width=\linewidth]{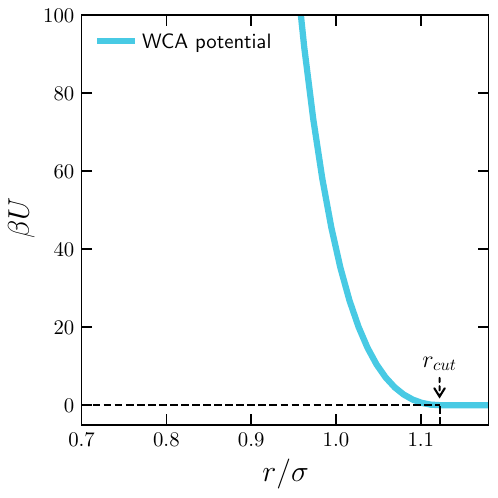}
\end{center}
\vspace{-0.5cm}
\caption{Weeks-Chandler-Andersen potential $\beta U(r)$ as a function of interparticle distance $r$.
The interaction cutoff distance is $r_{cut} = 2^{1/6}~\sigma$.}
\label{fig:WCA_potential}
\end{figure}

To describe the translational motion of individual colloidal particles within the solvent, we employ the overdamped Langevin equation  
\begin{equation} \label{eq:integration_scheme}
    \bm{r}_i(t + \delta t) = \bm{r}_i(t) + \frac{D}{k_B T} \bm{F}_i(t)\delta t + \delta \bm{r}_i^G,
\end{equation}
where $D = 0.0083~\sigma^2/\tau$ represents the diffusion coefficient~\cite{thorneywork2016gaussian}, $\bm{F}_i(t) = -\sum_{j\neq i} \partial U(r_{ij})/\partial \bm{r}_i$ denotes the force acting on particle $i$ due to interparticle interactions, which can be complemented with the rotational force $\bm{F}_{rot}(\bm{r}_i)$. Here, $r_{ij}=|\bm{r}_i-\bm{r}_j|$ represents the distance between particle $i$ and $j$, and $\delta \bm{r}^G$ is the random displacement,  with its components independently sampled from a Gaussian distribution with variance $\sqrt{2 D \delta t}$ and zero mean.
We employ the Weeks-Chandler-Andersen (WCA) potential  to model the particle interactions 
\begin{equation} \label{eq:WCA_potential}
    U(r_{ij}) = 
    \begin{cases} 
      4 \epsilon \left( \left( \frac{\sigma}{r} \right)^{12} - \left( \frac{\sigma}{r} \right)^6 \right) + \epsilon & r < r_{\text{cut}} \\
      0 & r \geq r_{\text{cut}}
   \end{cases},
\end{equation}
where $r_{\text{cut}} = 2^{1/6} \sigma$. 
We fix the temperature $k_B T / \epsilon = 0.025$ to mimic hard spheres~\cite{coli2021artificial}, as also illustrated in Fig.~\ref{fig:WCA_potential}. 

We perform Brownian dynamics simulations using Eq.~\ref{eq:integration_scheme} for various combinations of  initial radii, misorientation angles, and packing fractions, to explore different regions of the parameter  space. The initial radii considered are $R_0/\sigma = 10, 15, 18, 20, 25, 30, 35, 40, 50$, and $60$. The initial misorientation angles range from $\theta_0 = 4, 7, 10, 16, 22$, to  $28\degree$, and we explore a packing fraction $\eta=\pi \sigma^2 N/ 4A_{box} = 0.68$, with $N$ the number of particles and $A_{box}$ the area of the simulation box. We consider system sizes of  $2900 \lesssim N \lesssim 21500$ particles within a box of area  $3350 \lesssim A_{box}/\sigma^2 \lesssim 24800$, depending on the size of the initial radius $R_0$.
Finally, we set the time $t = 0$  when the rotational force responsible for forming  the grain is turned off, and we define $t = t_f$ as the  time required for the grain to collapse, i.e. the point at which the GBL entirely loses its loop characteristics, when $R \theta(t) < 3 \sigma / \pi$~\cite{lavergne2018dislocation}. 

\section{Experimental details}

\subsection{Colloidal system}

Two-dimensional monolayer colloidal crystals were created using melamine-formaldehyde microspheres (microParticles GmbH) with a diameter of $\sigma = 2.82\mu$m.  These spheres were dispersed in a  ethanol/water mixture (20/80 v/v\%), resulting in hard-sphere interactions as detailed in \cite{thorneywork2017two}.  The suspension was then introduced into a $200\mu$m thick quartz cell (Hellma Analytics).  Due to the significant density mismatch between the particles and the solvent, the spheres rapidly sediment, forming a quasi-2D monolayer with minimal out-of-plane fluctuations.  By using a sufficiently high area fraction ($\eta \approx 0.74-0.76$) a colloidal crystal formed.  The samples were then left to coarsen for approximately one week, resulting in large crystal domains that span the entire field of view.

\subsection{GBL creation and imaging}
GBLs were created using a holographic optical manipulation technique analogous to that reported in \cite{lavergne2018dislocation}.  A spatial light modulator (Meadowlark) was employed to generate a defocused optical vortex, which was used to rotate a circular region of the 2D colloidal crystal, creating a controlled misorientation.  This process yielded loop-shaped GBs with a typical diameter of approximately 70 lattice spacings.  Once the target misorientation was achieved, the optical vortex was deactivated, allowing the created domain to relax and shrink free from external forces.  The GBLs were created within large, single-domain regions of the crystal to ensure their isolation from other GBs.  Bright-field video microscopy was used to observe the GBL shrinkage at the single-particle level. Particle positions were extracted using an established particle tracking routine~\cite{crocker1996methods}.

\section{Analysis details}

\subsection{Detection of the grain}
\begin{figure*}[t!]
\begin{center}
\includegraphics[width=1.0\linewidth]{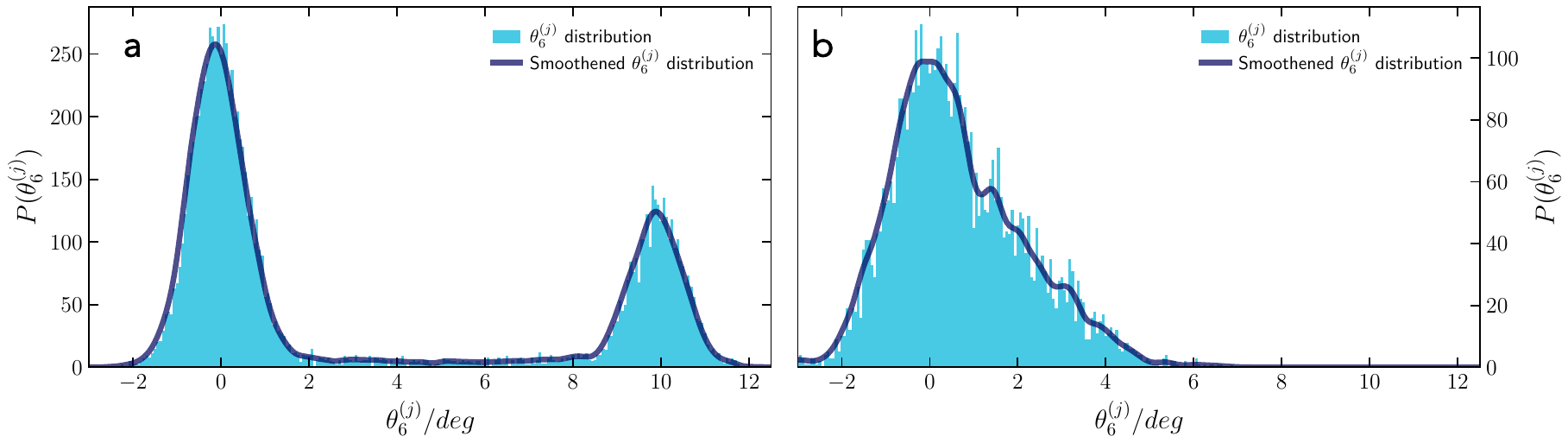}
\end{center}
\vspace{-0.5cm}
\caption{The probability distribution of  local orientations of particles $P(\theta^{(j)}_6)$ in  a 2D colloidal crystal with a GBL (a) when bulk and grain particles have  fully distinct local orientations, and (b) when they possess similar local orientations.  
The data corresponds to (a) $R_0 = 35\sigma$, $\theta_0 = 10\degree$, $\eta=0.68$ and (b) $R_0 = 25\sigma$, $\theta_0 = 4\degree$, $\eta=0.65$.}
\label{fig:histograms}
\end{figure*}

To detect the GBL, we calculate the local orientation $\theta_6^{(j)}$ for each particle $j$, which is determined through the hexatic order parameter 
\begin{equation} \label{eq:bond_order_parameter}
    \tilde{\Psi}_6^{(j)} = \frac{1}{N_j} \sum_k e^{i6 \alpha_{jk}},
\end{equation}
where $N_j$ represents the total number of neighbours of particle $j$, and $\alpha_{jk}$ is the angle between $\bm{r}_{jk}=\bm{r}_j-\bm{r}_k$ and the $x$-axis.  The sum in Eq.~\ref{eq:bond_order_parameter} runs over all neighbouring particles $k$ within a radial distance of $r < 1.4~r_{\text{cut}}$.
The choice of this limit is determined empirically to include the nearest neighbours while excluding all particles beyond the first shell. Since this definition is susceptible to noise, we locally average the hexatic order parameter for each particle $j$ over all neighbouring particles $\Psi_6^{(j)} = (\tilde{\Psi}_6^{(j)}+\sum_k \tilde{\Psi}_6^{(k)})/(N_j+1)$. 
Subsequently, we  calculate the local orientation of particle $j$ using 
\begin{equation} \label{eq:local_orientation}
    \theta_6^{(j)} = \arg \left( \Psi_6^{(j)} \right)/6.
\end{equation}
The  probability distribution of  local orientations  of particles $P(\theta_6^{(j)})$ typically reveals two distinctive peaks, arising from the clear difference between particles belonging to the grain and  those in the bulk, as shown in Fig.~\ref{fig:histograms}a. At the GBL, there is a sharp transition between these two domains, accounting for  the lower values in between the peaks. The histogram is  smoothed using a series of Gaussian filters, after which we determine the angle with the highest count in the peak that is not centered around $\theta^{(j)}_6 = 0\degree$. This angle is taken as the misorientation angle $\theta(t)$ of the grain at time $t$. 

All particles with a local orientation $1.5\theta(t)\geq \theta_6^{(j)}\geq 0.33\theta(t)$ are selected as potential grain particle candidates. These threshold values have been determined empirically  to yield the best results for grain detection across  the range of  misorientation angles studied here.

\begin{figure*}[t!]
\begin{center}
\includegraphics[width=1.0\linewidth]{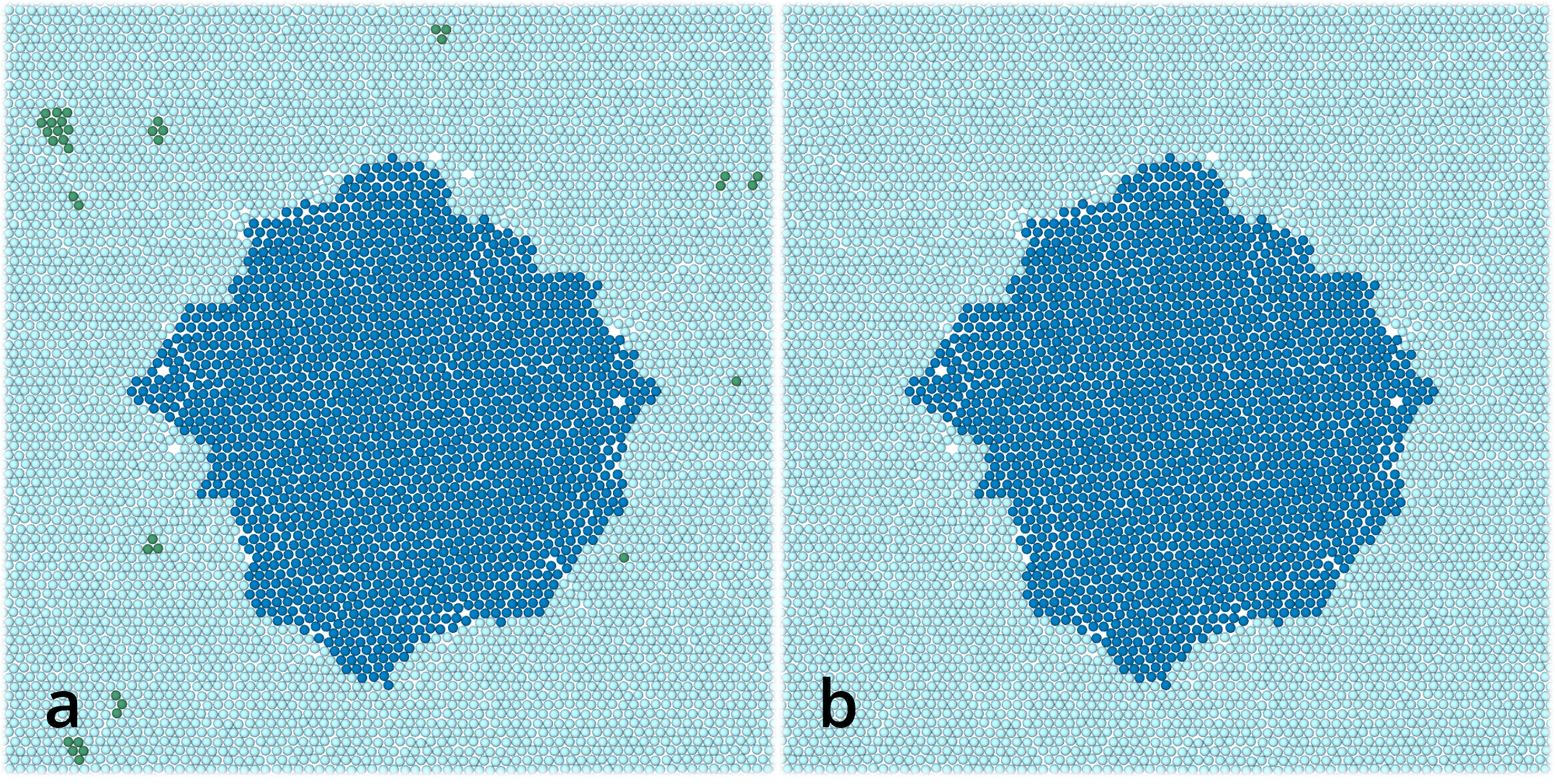}
\end{center}
\vspace{-0.5cm}
\caption{a) Grain detection is performed by selecting all particles with a local orientation $\theta^{(j)}_6 > 0.33~\theta(t)$. However, for small misorientation angles, this method may inadvertently select  some bulk particles, which are colored in green. b) The grain detection is improved by  
applying a breadth-first search algorithm to exclude all particles that are not directly connected to the center of the grain.}
\label{fig:grain_detection}
\end{figure*}

However, for small misorientation angles, there may be a small overlap between the two peaks, as shown in Fig.~\ref{fig:histograms}b. 
When a distinct peak is not observable, we define $\theta(t)$ as the angle at which the smoothed $\theta^{(j)}_6$ distribution has a slope closest to zero.
Furthermore, selecting all particles with a local orientation $\theta^{(j)}_6 > 0.33~\theta(t)$ may also include particles belonging to the bulk crystal that are not connected to the grain, see Fig.~\ref{fig:grain_detection}. To exclude these particles, we first identify the most central particle belonging to the grain and then use a breadth-first search~\cite{BreadthFirstSearch} to find all particles directly connected to the central grain.
Once the number of particles belonging to the grain $N_{grain}$ are determined, the area of the grain is calculated as $A(t)=N_{\text{grain}}(t)\sigma^2\pi/ 4\eta$, and its radius as $R(t)=\sqrt{N_{grain}(t)/\pi} a$, with $a$ the lattice spacing~\cite{lavergne2019shrinkage}. 
Finally, we remark that, during the simulation run, the GBL can be mistakenly identified. These (few) isolated  data points are excluded from the following analysis, as they do not represent the system correctly.

\subsection{Shrinkage mechanisms}
\begin{figure*}[t!]
\begin{center}
\includegraphics[width=1.0\linewidth]{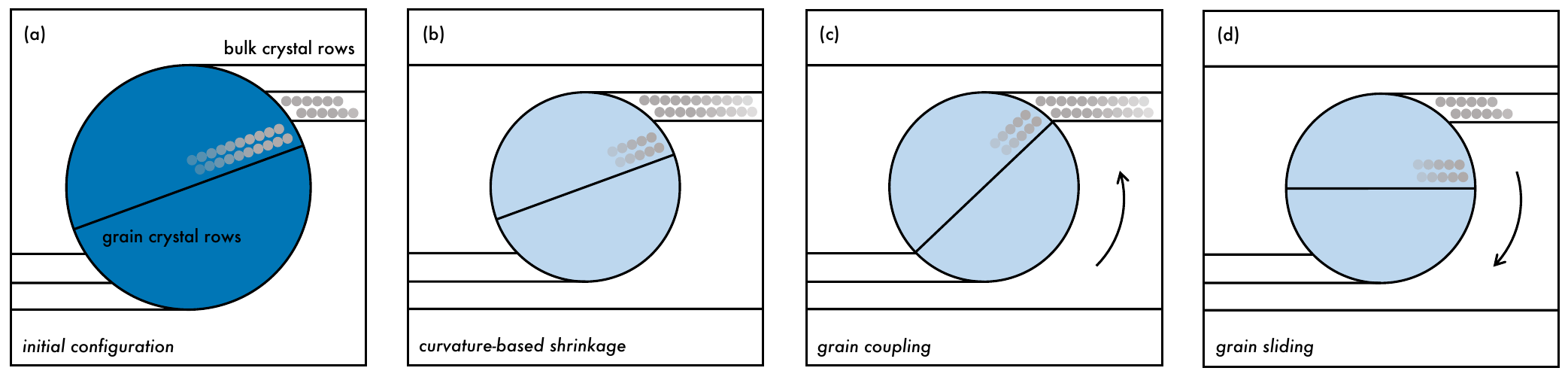}
\end{center}
\vspace{-0.5cm}
\caption{Schematics of the grain shrinkage mechanisms. (a) Initial configuration showing the bicrystalline system with the bulk (horizontal lines) and the grain crystal (tilted line) rows. (b) Curvature-based grain shrinkage: the misorientation angle remains unchanged while  particles at the GB adopt positions that align with the bulk crystal rows. (c) Grain coupling: the misorientation angle increases as particles at the GB maintain their connectivity to both the host and bulk crystal rows during the shrinkage process. (d) Grain sliding: the misorientation angle decreases during the shrinkage process.
}
\label{fig:scheme_shrinking_mechanisms}
\end{figure*}

\begin{figure*}[t!]
\begin{center}
\includegraphics[width=1.0\linewidth]{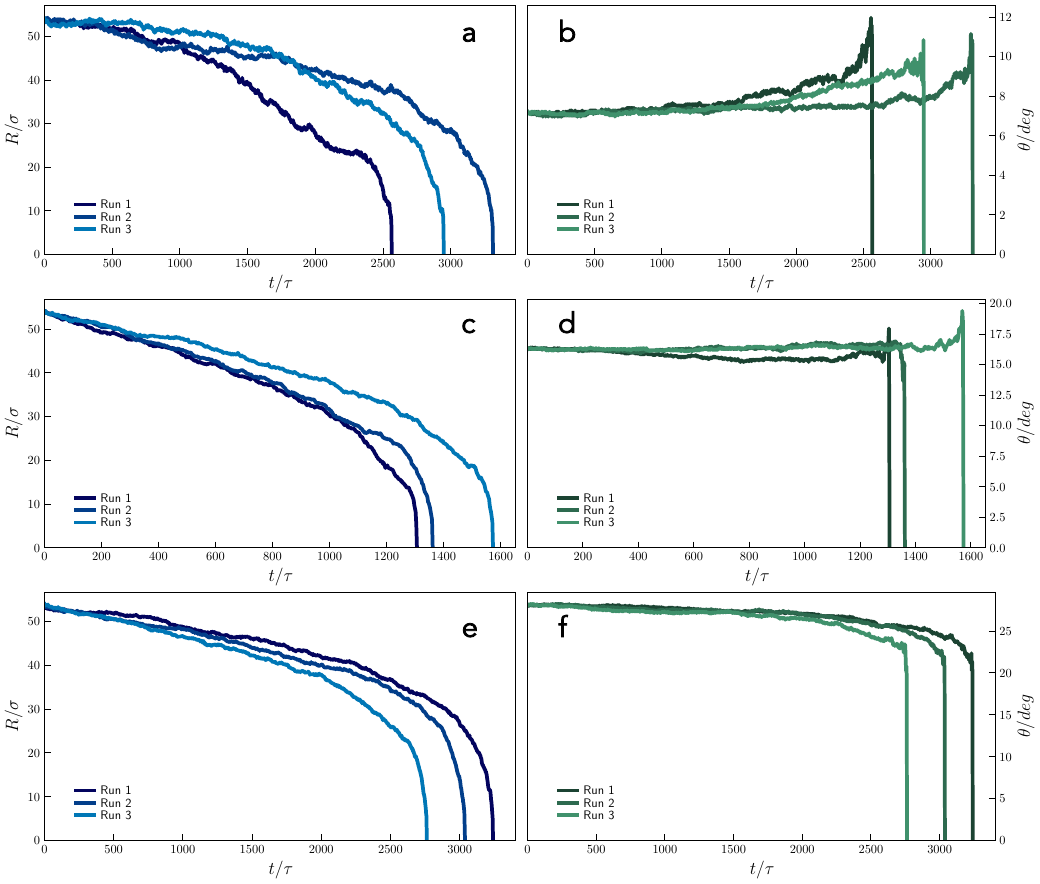}
\end{center}
\vspace{-0.5cm}
\caption{(Left) Evolution of grain radii $R(t)$ and (right) misorientation angles $\theta(t)$ as a function of simulation time $t/\tau$ for initial misorientation angles (a,b) $\theta_0 \approx 28\degree$, (c,d) $\theta_0 \approx 16\degree$ and (e,f) $\theta_0 \approx 7\degree$. For each configuration three runs are shown, as indicated in the legend. All systems start with an  initial radius $R_0 \approx 50\sigma$.
}
\label{fig:shrinking_mechanisms}
\end{figure*}

\begin{figure*}[t!]
\begin{center}
\includegraphics[width=1.0\linewidth]{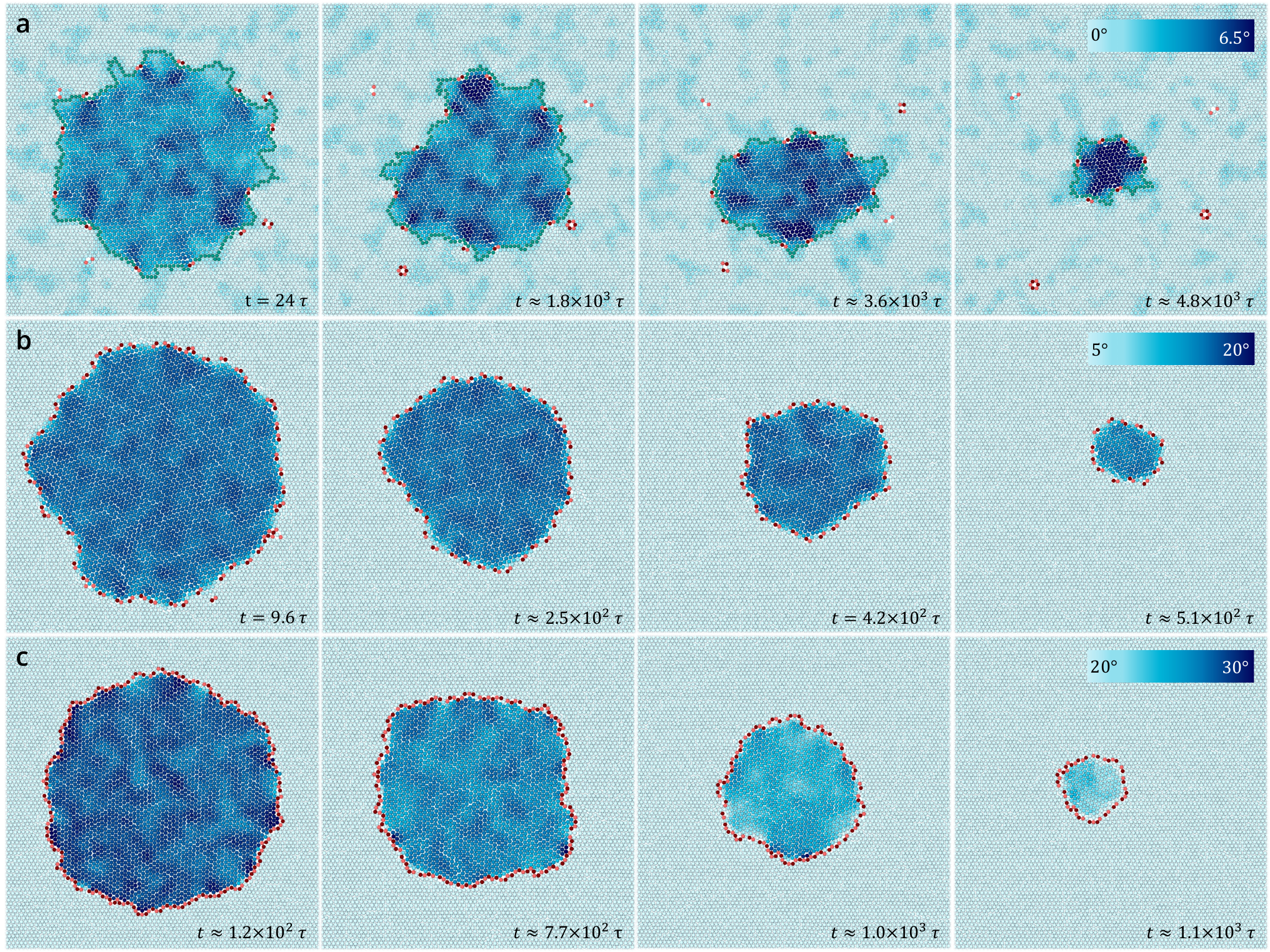}
\end{center}
\vspace{-0.5cm}
\caption{Representative time-lapse simulation snapshots demonstrating the shrinkage of GBLs for (a) grain coupling, (b) curvature-based shrinkage and (c) grain sliding. The color of each particle corresponds to its local orientation $\theta_6^{(j)}$,  as indicated by the color bar in each series of snapshots. To enhance  visual clarity, (b) and (c) are color-coded based on $\theta_6^{(j)} > 0\degree$, while bulk particles still have local orientations $\theta_6^{(j)} \approx 0\degree$. Dislocations are represented as pink/red particle pairs. In (a) the GBL is highlighted in green, whereas in (b) and (c), the GBL is roughly within the area enclosed by the dislocations. The simulation time $t/\tau$ is indicated in the bottom right corner of each panel.}
\label{fig:shrinking_snapshots}
\end{figure*}

A summarizing schematic and exemplary trajectories for the main shrinking mechanisms are depicted in Fig.~\ref{fig:scheme_shrinking_mechanisms} and Fig.~\ref{fig:shrinking_mechanisms}, respectively. Similarly, Fig.~\ref{fig:shrinking_snapshots} shows representative time-lapse simulation snapshots.


\begin{figure*}[ht!]
\begin{center}
\includegraphics[width=0.8\linewidth]{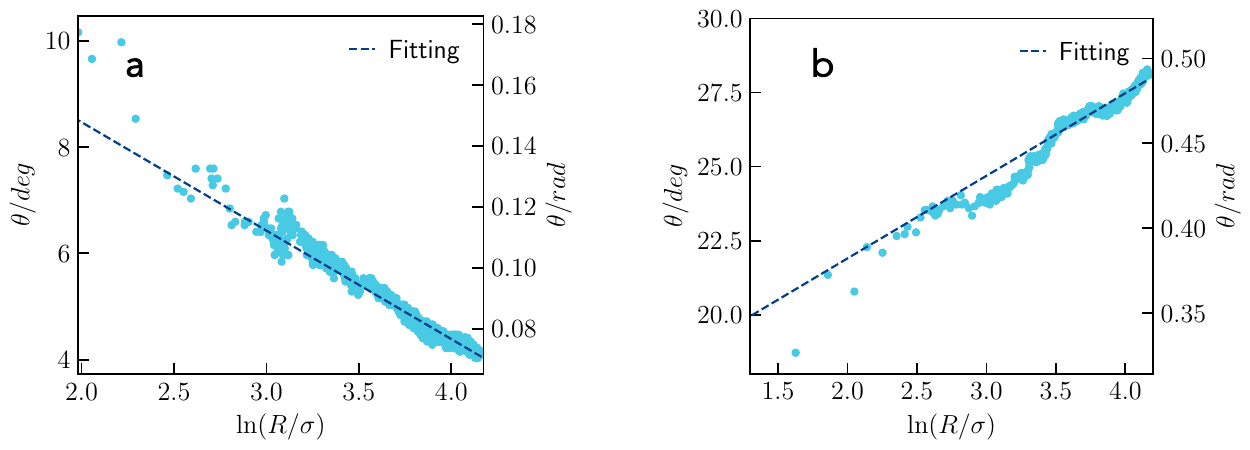}
\end{center}
\vspace{-0.5cm}
\caption{Linear fit of the misorientation angle $\theta(t)$ as a function of the logarithm of the grain radius $R(t)$ to determine the dominant shrinking mechanism for a system with a given initial radius $R_0$ and misorientation angle $\theta_0$. The slope of the fit corresponds to the shear coupling parameter $\beta$. The data is for an initial radius $R_0 \approx 60\sigma$ and initial misorientation angles (a) $\theta_0 = 4\degree$, and (b) $\theta_0 = 28\degree$, corresponding to grain coupling,  and grain sliding, respectively.}
\label{fig:statespace_fitting}
\end{figure*}

To distinguish the dominant shrinking mechanism in a simulation run, we make use of the shear coupling factor $\beta$, defined as the ratio between the tangential and the normal GB velocities,
\begin{equation}
    \beta=\frac{v_\parallel}{v_\perp}=-\frac{d\theta/rad}{d \ln{R/\sigma}}.
\end{equation}
We compute $\beta$ for all simulation state points and experiments by fitting a linear function to the decay of misorientation angle in radians as a function of the logarithm of the GBL radius for $t<0.7~t_f$. The value of $\beta$ is used to assign color shades to the points in the state diagram in Fig.~1 in the main text. Consequently,  different colors are used to indicate the prevailing type of shrinkage mechanism for each state point. The data in the state diagram represents the result of averaging over at least three simulation runs.  
Fig.~\ref{fig:statespace_fitting} presents two example fittings representing  the coupling and sliding shrinking mechanisms, respectively.

\subsection{Perimeter of the grain}
\begin{figure*}[h]
\begin{center}
\includegraphics[width=0.35\linewidth]{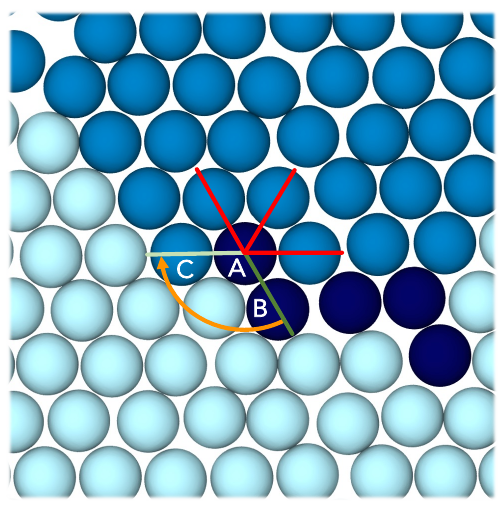}
\end{center}
\vspace{-0.5cm}
\caption{Schematic representation of the algorithm used to determine the next particle belonging to the perimeter of the grain. Dark blue particles have already been identified to lie on the perimeter, light blue particles lie outside the grain, and blue particles are part of the grain.
We begin with particle A and first identify the previously determined perimeter particle (B), which is highlighted by a dark green line. Then, we select the next grain particle belonging to the perimeter (C) as the one  with the smallest angle in the clockwise direction (light green line and orange arrow) among those belonging to the GBL. The red lines aid visualizing the angles between B and  potential next perimeter particles. However, these particles are not selected as they form larger angles in the clockwise direction with B compared to C.
}
\label{fig:grain_boundary_detection}
\end{figure*}

\begin{figure*}[h]
\begin{center}
\includegraphics[width=1.0\linewidth]{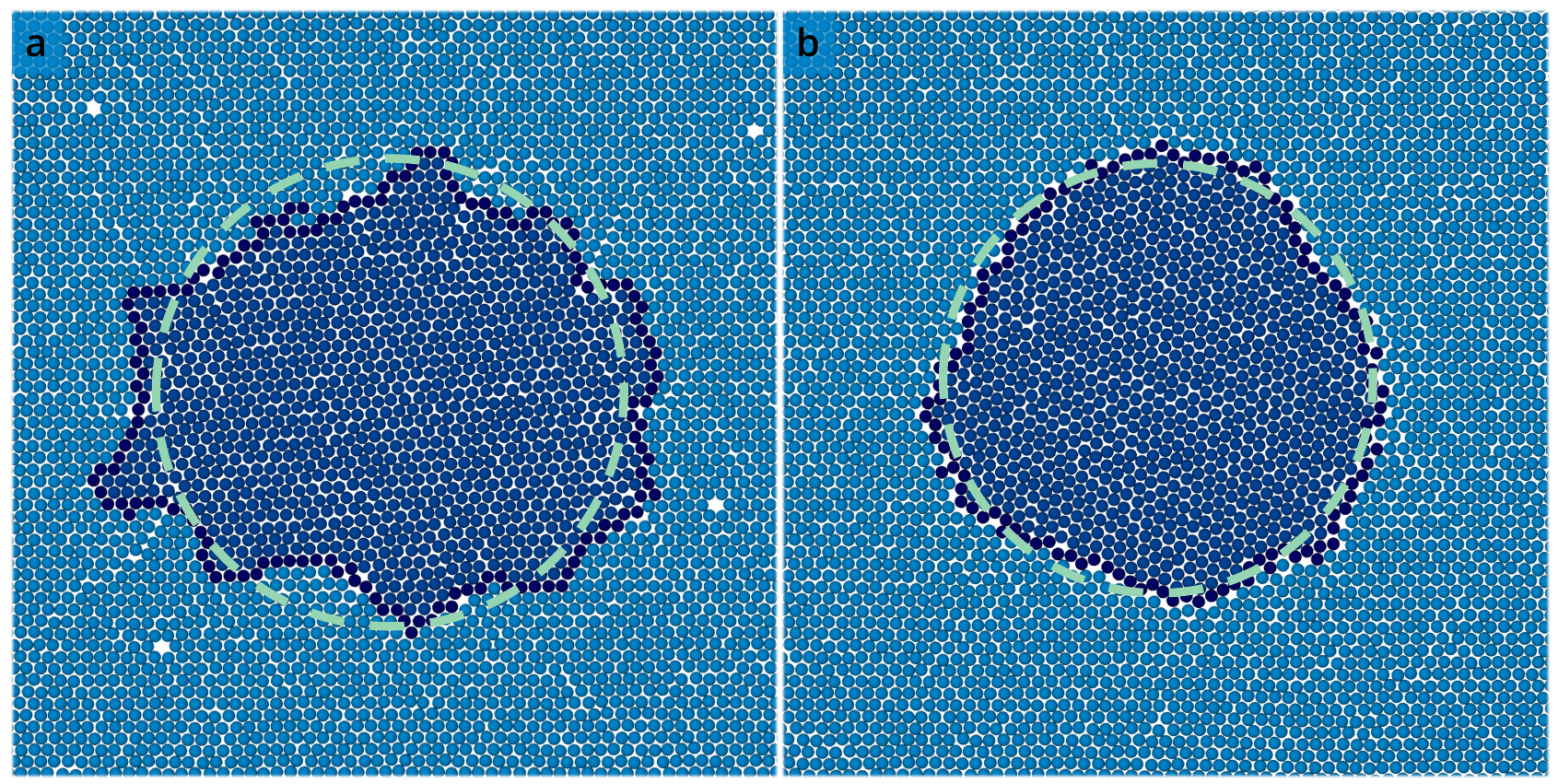}
\end{center}
\vspace{-0.5cm}
\caption{Schematics illustrating the perimeter length $L_{GBL}$ of the GBL (dark blue particles) as compared to a circle (dashed green line) of equal  area as the grain  for (a) $\theta_0 = 4\degree$ and (b) $\theta_0 = 28\degree$. The grain particles are colored blue and the particles belonging to the bulk crystal light blue. 
}
\label{fig:GBL_length_schematic}
\end{figure*}

Once we have identified the particles belonging to the grain, we can determine its perimeter length $L_{GBL}$. To this aim, we first locate the grain particle with the lowest y-coordinate and take that as our starting point, which is by default part of the perimeter. We then proceed by looping clockwise around the grain, as shown in Fig.~\ref{fig:grain_boundary_detection}, where the algorithm is schematized.
Once all the particles belonging to the perimeter are detected, the perimeter length can be computed by summing all the interparticle distances encountered while looping around the grain. 
Simulation snapshots illustrating the perimeter length in comparison to a circle with  the same area as the grain are reported in Fig.~\ref{fig:GBL_length_schematic}.

\subsection{Dislocation density}
\begin{figure*}[h]
\begin{center}
\includegraphics[width=0.5\linewidth]{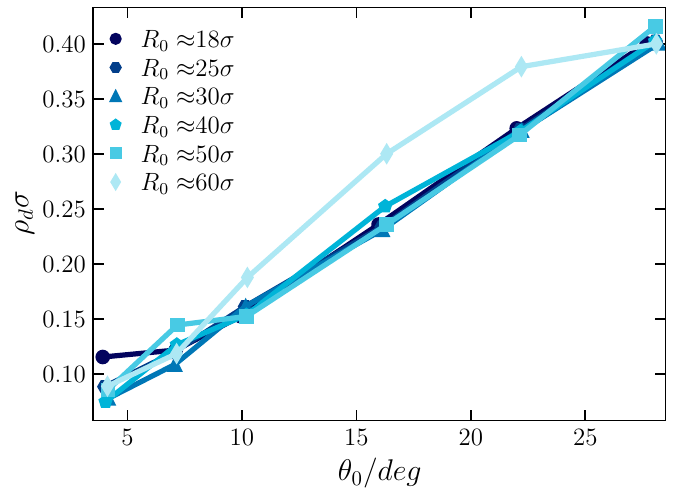}
\end{center}
\vspace{-0.5cm}
\caption{Dislocation density $\rho_d\sigma$ along the GBL as a function of the initial misorientation angle $\theta_0$ for various initial grain radii $R_0$, as indicated in the legend.}
\label{fig:dislocation_density}
\end{figure*}

To determine the number of dislocations $n_d$ in a simulation snapshot, we first construct the Voronoi tessellation~\cite{rycroft2009voro++} of all particles. 
Dislocations are identified as particle pairs that have five or seven neighbors. Occasional particles with eight neighbors have always two adjacent particles with five neighbors, and can be considered part of two dislocations. Similarly, particles with four neighbours have two adjacent particles with seven neighbours and can also be regarded as two dislocations. 
We only count  dislocation pairs  located within a $3\sigma$ distance from the grain boundary. This threshold distance is empirically determined to include all relevant dislocations while excluding any leftover point defects in the bulk crystal. To calculate the dislocation density, we divide the number of dislocations  by the perimeter length:
\begin{equation}
    \rho_d = n_d / L_{GBL}.
\end{equation}
Fig.~\ref{fig:dislocation_density} reports $\rho_d$ as a function of the initial misorientation angle $\theta_0$ for various initial grain radii $R_0$. We observe that $\rho_d$ does not exhibit a dependence on $R_0$, instead, it monotonously increases with $\theta_0$~\cite{hirth1983theory}.

\subsection{Imperfect coupling and dislocation preservation}

At small misorientations the shear coupling parameter is expected to follow the relation $\beta_\mathrm{geom}=2\tan(\theta/2)$ when the GBL is perfectly coupled with the crystal and the shrinking is governed only by geometry~\cite{cahn2004unified}. However, from Fig.~1c we observe that the measured shear coupling parameter is always smaller than $\beta_\mathrm{geom}$. We explain this from the fact that the shrinking mechanism seems to follow an imperfect coupling scenario, with $\beta=f_c\beta_\mathrm{geom}$ modulated by a factor $f_c$. To obtain $f_c$, we fit the shrinking behavior of $\theta(R(t))$ with Eq.~1 from the main text. 

\begin{figure}[h]
\begin{center}
\includegraphics[width=0.9\linewidth]{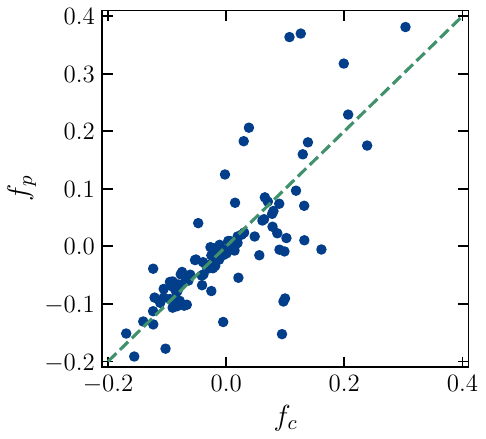}
\end{center}
\vspace{-0.5cm}
\caption{Comparison between the fraction of dislocations preserved $f_p$ and the parameter $f_c$ for all analyzed systems. Dashed line shows the $f_p=f_c$. }
\label{fig:f_comparison}
\end{figure}

The imperfect coupling behavior can also be linked microscopically to the preservation of dislocations during shrinkage. As shown in the main text, the 
dislocation content of the GBL can 
be described by a linear combination
of two limit cases. In the first case, where the grain is perfectly coupled, shrinkage occurs without dislocation annihilation, so the number of dislocations remains constant, that is $n_d(t)=n_{d,0}$, with $n_{d,0}$ the initial number of dislocations. In the second case, no grain rotation occurs during the shrinkage, leading to dislocation annihilation to maintain a constant dislocations number density, so that the number of dislocations is described by $n_d(t)=n_{d,0} R(t)/R_0$. A linear combination of these two limiting behaviors can be expressed as:
\begin{equation}
  n_d(t)=f_p n_{d,0}+(1-f_p) n_{d,0} \frac{R(t)}{R_0}
\end{equation}
where $f_p$ is the fraction of dislocations being preserved during the shrinkage process. Rewriting this as a dislocation number density:
\begin{equation}
    \rho_d(t)=f_p \frac{n_{d,0}}{L_{GBL}(t)}+(1-f_p)\frac{n_{d,0}}{L_{GBL}(t)}\frac{R(t)}{R_0}.
\end{equation}
Finally, using the relation $n_{d,0}=\rho_{d,0}L_{GBL,0}$ and $L_{GBL}(t)/L_{GBL,0}=R(t)/R_0$, we arrive at Eq.~2 of the main text:
\begin{equation}
    \rho_d(R)  = f_p \rho_{d,0} R_0/R +(1-f_p) \rho_{d,0}.
\end{equation}
 Fitting Eqs.~1 and 2 of the main text to the measured shrinkage trajectories reveals a strong correlation between $f_c$ and $f_p$ (Fig.~\ref{fig:f_comparison}), with $f_c \approx f_p$. This correspondence is not merely observational, given that the approximation $\rho_d a\approx \theta$ holds,  Eqs.~1 and 2 are mathematically equivalent, implying that $f_c=f_p$.

\subsection{Window-averaged shrinkage rate}

To clearly illustrate the change in shrinkage kinetics at different loop sizes, we define a window-averaged shrinkage rate as $\mathcal{M}_w = A(t)/2\pi(t_f-t)$ over the time window $[t,t_f]$, where $t_f$ is the time at which the loop is completely annihilated. This approach effectively discards early-stage shrinkage data within the time window $[0,t]$. Plotting $\mathcal{M}_w$ for various $R$ (i.e. different time windows), a significant shift in the shrinkage behavior becomes evident below a critical loop size $R \lesssim 28\sigma$. While in the large-loop regime, the shrinkage kinetics approaches a constant rate $\mathcal{M}_w= M^\ast$, the small-loop regime is characterized by non-linear and  significantly faster dynamics, making $\mathcal{M}_w$ strongly dependent on the loop size. 

\subsection{Initial shrinkage rate}

\begin{figure}[th!]
\begin{center}
\includegraphics[width=0.9\linewidth]{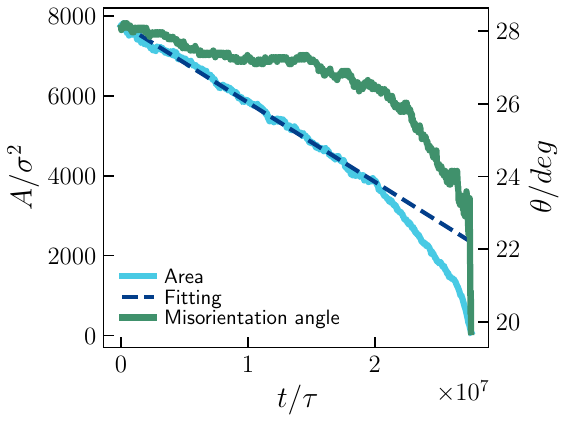}
\end{center}
\vspace{-0.5cm}
\caption{Fitting of a linear function (dashed blue line) to the time evolution of the area $A(t)$ (light blue solid line) to extract the mobility $\mathcal{M}_0$ using a fit of $A(t)$ up to $t = 0.7t_f$. The left axis displays the area $A(t)$, while the right axis shows the misorientation angle $\theta(t)$ (green solid line).}
\label{fig:mobility_fitting}
\end{figure}

The initial shrinkage rate, $\mathcal{M}_0=-\langle dA/dt\rangle /2 \pi $, is determined by analyzing the area decay during the \emph{initial} stage of GBL shrinkage. Specifically, we perform a linear fit on the area evolution $A(t)/\sigma^2 = a + b t/\tau$, as shown in Fig.~\ref{fig:mobility_fitting}, from which we extract $\mathcal{M}_0 = -b \sigma^2/\tau$. 
This fit is restricted to the initial shrinkage stage, $t<0.7~t_f$, in order to isolate the linear shrinkage regime and avoid the influence of non-linear kinetics present at smaller loop sizes. 
Consequently, for large grain radii $R_0$, $\mathcal{M}_0$ is constant and  provides a direct measure for the reduced mobility $\mathcal{M}_0=M^\ast$.

\begin{figure}[ht!]
\begin{center}
\includegraphics[width=0.9\linewidth]{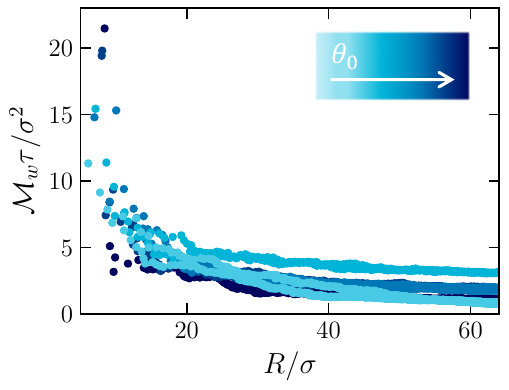}
\end{center}
\vspace{-0.5cm}
\caption{Window-averaged shrinkage rate $\mathcal{M}_w$ for $R_0=60\sigma$, varying $\theta_0$ and different runs. 
}
\label{fig:shrinkdt}
\end{figure}

\clearpage
\newpage

\bibliographystyle{apsrev4-1}
\bibliography{grain}

\end{document}